\documentclass[a4paper,11pt]{article}
\usepackage{amsmath}

\begin{document}

\newcommand{\beq}{\begin{equation}}
\newcommand{\eeq}{\end{equation}}
\newcommand{\beqa}{\begin{eqnarray}}
\newcommand{\eeqa}{\end{eqnarray}}

%% Handy Dirac notation.
\def\ket#1{|\,#1\,\rangle}
\def\bra#1{\langle\, #1\,|}
\def\braket#1#2{\langle\, #1\,|\,#2\,\rangle}
\def\ketbra#1#2{\ket{#1}\bra{#2}}
\def\identity{\leavevmode\hbox{\small1\kern-3.8pt\normalsize1}}  %Makes an "open 1"

%% Formatting instructions
\def\lineacrosspage{\hbox to \hsize{\hfill\rule[5pt]{2.5cm}{0.5pt}\hfill}}
\def\FLAG{ \par \medskip \noindent \emph{[This Section Not Yet Complete]} \par \medskip }

%% A bit of maths.
\def\set#1{\{ #1\}}
\def\Prob#1{\mbox{Prob}(#1)}
\def\modulus#1{\left| #1 \right|}
\newcommand{\QED}{\hspace*{\fill}\mbox{\rule[0pt]{1.5ex}{1.5ex}}}
\newcommand{\qed}{\mbox{\rule[0pt]{1.5ex}{1.5ex}}}
\newcommand{\half}{\mbox{$\textstyle \frac{1}{2}$} }
\def\indicator#1{\left\{ \phantom{\big|} #1 \phantom{\big|}\right\}}

%% Some environments
\newtheorem{theorem}{Theorem}
\newtheorem{definition}{Definition}
\newtheorem{lemma}{Lemma}
\newtheorem{example}{Example}
\newtheorem{property}{Property}
\newtheorem{proposition}{Proposition}
\newtheorem{corollary}{Corollary}
\newtheorem{conjecture}{Conjecture}

%% Making title, etc.
\title{On the Role of Hadamard Gates\\ in Quantum Circuits\\                         }%  \phantom{|}\small{(DRAFT)}}
\author{D.~J.~Shepherd\footnote{shepherd@compsci.bristol.ac.uk,  dan.shepherd@cesg.gsi.gov.uk}\\
\small{\it University of Bristol,  Department of Computer Science}\\
}
\date{March 23, 2006}
\maketitle

%% ABSTRACT
\abstract{
We study a reduced quantum circuit computation paradigm in which the only allowable gates either permute the computational basis states or else apply a ``global Hadamard operation'', \emph{i.e.} apply a Hadamard operation to every qubit simultaneously.
In this model, we discuss complexity bounds (lower-bounding the number of global Hadamard operations) for common quantum algorithms~: we illustrate upper bounds for Shor's Algorithm, and prove lower bounds for Grover's Algorithm.  We also use our formalism to display a gate that is neither quantum-universal nor classically simulable, on the assumption that Integer Factoring is not in \textbf{BPP}.
} \vspace{1 cm} \normalsize

%% SECTION ONE : The Introduction
\section{Introduction}  \label{sect:intro}
A Quantum Circuit (or Quantum Logic Network) is usually presented as being composed both of \emph{wires} that carry qubits and \emph{gates} that tap those wires to modify the qubits they carry, \cite{NandC}.
In section \ref{sect:notate} we specify the notation used to describe computation with quantum circuits, and specify exactly which features we shall be allowing within the kinds of circuits we wish to consider.

The main focus is to enquire about the difference it makes if we limit to using `classical' gates (ones which preserve the set of computational basis states) and `global Hadamard transforms' (where a Hadamard gate is applied once to each qubit.)  We will show in section \ref{sect:universality} that our imposed limitations do not limit computational power in any real sense, and in subsequent sections will discuss what algorithms and algorithm-primitives tend to look like within this model.  This model is closely related to the complexity class \textbf{Fourier Hierarchy}, defined at \cite{web:zoo}.
The motivation comes not from physical considerations pertinent to the task of fabricating a quantum information processor, but from the desire to analyse a fairly natural-looking measure of circuit complexity that is not apparent within the standard model, {\it viz} the number of these global Hadamard transformations needed.  The reason for requiring that the Hadamard operations be applied on every qubit is that we're not necessarily thinking of \emph{actively} applying them, so much as just \emph{passively} changing what is meant by the computational basis, and therefore changing the interpretation of future gates or measurements.  More will be said on this in Section \ref{sect:Qdepth}.

Section \ref{sect:Shor} will discuss the idea of replacing the Fourier Transform in Shor's Algorithm with the kind of transform that is simplest within our limited model, and look at how this affects the gate-complexity of the algorithm.  The concept of \emph{Order Finding}, on which Shor's Algorithm is based, turns out to be a very natural primitive within the present context.

In Section \ref{sect:Grover}, we extend a result of \cite{0312213} to show the trade-offs in different complexity measures relevant to Grover's Algorithm.  This is undertaken in a similar spirit to the work of \cite{Zalka}, showing why Grover's algorithm is essentially non-parallelisable.

%% SECTION TWO :  Notation
\section{Notation and Terminology}  \label{sect:notate}
A circuit incorporates a finite number of wires, which run right through the circuit, not terminating prematurely (\emph{i.e.} no `adaptive' measurements.)  The \emph{width} of a circuit is taken to be the number of such qubit wires used.

The input to the circuit is taken to be a (classical description of the) state in which the wires are initialised, usually a simple computational basis state.  

Each wire, {\sl i.e.} qubit, codes quantum data with respect to a tensor component ${\bf C}^2$ within the full space ${\bf C}^{2^n}$ that is associated with the totality of $n$ qubits, and these component spaces are given a \emph{computational basis} ($Z$) and a \emph{Hadamard basis} ($X$), both of which are orthonormal~:
\beqa
  \mbox{Z-basis} &:=& \Big\{~ \ket0,~~   \ket1 ~\Big\}, \nonumber \\
  \mbox{X-basis} &:=& \left\{~ \begin{array}{l} \ket+ ~:=~ H\ket0 ~:=~ { \frac{\ket0+\ket1}{\sqrt2} }, \\
                                   \ket- ~:=~ H\ket1 ~:=~ { \frac{\ket0-\ket1}{\sqrt2} } 
                  \end{array} ~\right\}.
\eeqa

The data resident on a circuit may equally well be described by a trace-1 Hermitian density operator, which will be of rank 1 if the state is pure.  We will use an algebra of Pauli symbols 
\beq
  \def\tmatrix#1#2#3#4{\left(\begin{array}{cc}#1&#2\\#3&#4\end{array}\right)}
 I = \tmatrix1001, \quad X = \tmatrix0110, \quad Y = \tmatrix0{-i}i0, \quad Z = \tmatrix100{-1},
\eeq
to describe such density operators explicitly, where necessary.  Subscripts on these symbols will distinguish between different qubits.
The same Pauli symbols may be used to describe unitary transforms (conjugations of the density operator by unitary actions,) so for example the \emph{C-Not gate} may be written
\beq
  \Lambda_c( X_t ) ~=~ \half( 1 + Z_c + X_t - Z_c \cdot X_t ),
\eeq
where $c$ and $t$ label the \emph{control} and \emph{target} qubits respectively.
More generally, the \emph{generalised Toffoli gate} may be written
\beq  \label{eqn:GenToffoli}
  \Lambda_C( X_T ) ~=~ 1 - \left( 1 - \prod_{t \in T}X_t \right) \prod_{c \in C}\left( \frac{1-Z_c}{2} \right).
\eeq
In classical terms, these gates flip each of the target bits if each of the control bits is set to 1, and otherwise acts as the identity.  The two sets of qubits $C$ and $T$ must, of course, be disjoint, and $T$ must not be empty.  (To see that (\ref{eqn:GenToffoli}) is unitary, observe that it is the identity minus twice a product of commuting projectors, and hence geometrically a reflection.)  We call the number of such gates used the \emph{size} of the circuit, (other authors give similar -- though different -- definitions of \emph{size}, but this point will not be important in this discussion.)

As well as allowing for Generalised Toffoli gates (to be used an arbitrary number of times on arbitrary qubits within a circuit,) we also allow for a Hadamard map to be applied simultaneously to every qubit in the circuit~:
\beq
  H^\infty ~=~ \prod_k \left( \frac{X_k + Z_k}{\sqrt2} \right),
\eeq
where the product is taken over every index, without exception.
We use the expression \emph{quantum-depth} to count the number of such operations used within a circuit.

For the purposes of obtaining computational outcomes, projective single-qubit measurements in the computational basis may be applied to the quantum output of a circuit.  We will use the term \emph{output} to refer to the sublist of the measured qubits that carry the data in which we're interested, since in many circuit designs it happens that many wires don't output useful data.

To avoid unnecessary discussion at the bit-level, we will use the term \emph{register} to denote a list of wires, and \emph{output register} to denote those wires that carry useful data at the end of the circuit.  If a circuit has been designed to be useful with different inputs, then the term \emph{input register} will be used to refer to those wires that are initialised differently on different runs of the circuit, and the term \emph{ancilla register} will refer to those wires that are always initialised in the same way for each run.  If a circuit is \emph{not} designed to be used with different inputs, then the term \emph{ancilla register} is used instead to refer to the qubits that aren't in the \emph{output register}.

%% SECTION THREE :  Universality
\section{Universality}  \label{sect:universality}
It isn't too hard to see that there are constant-width and -depth circuits which can be used to implement \emph{local} Hadamard transforms within this model, with certain ancill\ae{} used as catalysts.  For example, the circuit
\beq
  \Lambda_{cd}(X_b) \cdot H^\infty \cdot \Lambda_{cd}(X_a) \cdot H^\infty \cdot \Lambda_{cd}(X_b)
\eeq
clearly has effects which are local to qubits $a,b,c,d$.  In particular, it maps states as follows~:
\beqa
  \ket{1-00}  &\mapsto&  \ket{1-++} \nonumber \\
  \ket{1-01}  &\mapsto&  \ket{1--+} \nonumber \\
  \ket{1-10}  &\mapsto&  \ket{1-+-} \nonumber \\
  \ket{1-11}  &\mapsto&  \ket{1---}. 
\eeqa
Therefore, we could take qubits $a,b,c$ to be an ancilla in the state $\ket{1-?}$, (the question-mark denoting a totally arbitrary qubit,) and use the circuit above, followed or preceded by a swap on qubits $c$ and $d$, to achieve a \emph{local} Hadamard transform on qubit $d$ while returning the other qubits to their former states.  (Strictly speaking, the data on qubit $c$ is not returned to its former state unless that state were fully mixed.  But if we never `care' about that qubit, then no problem arises.)  Notice also that the \emph{swap gate} can be constructed from three C-Not gates, which are amongst the Generalised Toffolis.
(It is interesting that an ancilla was necessary for this construction.)  

We should point out that although Toffoli gates and local Hadamard gates are insufficient for achieving all unitary transforms directly, they are well known to constitute a \emph{Universal Set} of computing gates, \cite{0301040}, and so the limitations that we have imposed should not be thought of as especially restrictive.  Yet one does indeed seem to end up with some limitations on computing power if it is not possible to initialise ancilla qubits with both $X$- and $Z$-basis elements.

%% SECTION FOUR :  Shor, and Small Quantum Depth
\section{Circuits with Small Quantum-Depth}  \label{sect:Shor}
\subsection{A Modification to Shor's Algorithm}
In this section, we will show how Shor's Algorithm may, for all practical purposes, be divided into two parts; the first of which performs some \emph{order-finding} operations, the second of which converts `order-finding data' into an actual problem answer, (and is called \emph{postprocessing}.)  We will show that each of these parts \emph{on their own} can be performed efficiently without using any Hadamard maps, though Hadamard maps are necessary within the interface.
(While the division of Shor's Algorithm into two parts is not in itself a novel concept, it is hoped that a detailed analysis of its implementation within this particular model has something useful to teach about its algorithmic complexity.)

\subsection{Order Finding}  \label{sect:OrderFinding}
For \emph{Order-Finding}, which is the first part of our algorithm, let there be an ancilla register whose contents will code for elements of some presentation of a cyclic group $G=\langle g \rangle$, when in the computational $Z$-basis, and which more generally codes for superpositions of group elements.  Let there be an input register that starts out with each qubit in the pure state $\ket+$.

Let $U$ be the map on the ancilla register that carries $\ket{x}$ to $\ket{x \cdot g}$, which (by hypothesis, and by universality of Toffoli gates for classical computation) can be constructed from a reasonable number of (generalised) Toffoli gates.  We shouldn't generally care how $U$ maps vectors that aren't in the span of the coding for $G$, (which necessarily exist whenever $|G|$ is not a power of 2,) so to keep matters simple we shall require that $U$ acts as the identity on computational basis vectors that don't code for elements of $G$.  Again, it is straightforward to find efficient Toffoli circuitry for doing this by simply taking classical circuitry for the same problem and making it reversible.

A. Kitaev made the observation in \cite{9511026} that simple circuitry will assist in the estimation of one of the eigenvalues of such a map, $U$.  Ignoring the eigenspace of eigenvalue~1, the remaining eigenspaces of the domain of $U$ will each be 1-dimensional, and of the form
\beq
  \ket{\lambda_\omega} ~~:=~~ |G|^{-1/2} ~\sum_{j=1}^{|G|} \omega^{-j} \ket{g^j},
\eeq
for $\omega$ a complex $|G|$th root of unity.

Then, if the ancilla register were to hold such an eigenvector, we could apply the gate $\Lambda_r( U_{\rm anc} )$ -- a gate likewise constructible from generalised Toffoli maps -- which implements $U$ on the ancilla controlled on the $Z$-basis setting of an input-register qubit $r$; and thus effect the transformation
\beq
  \ket+_r = \frac{ \ket0_r +       \ket1_r }{\sqrt2} ~~\mapsto~~ 
            \frac{ \ket0_r + \omega\ket1_r }{\sqrt2} = \frac{(1+\omega)}{2}\ket+_r + \frac{(1-\omega)}{2}\ket-_r.
\eeq
The classical bit issued from the measurement (in the Hadamard basis) of the register wire would therefore be biased as
$[\cos^2( \theta ) : \sin^2( \theta )]$, where $\omega =: \exp( 2i\theta )$.  If we would rather not make a Hadamard-basis measurement for classical postprocessing but would instead process the data within the quantum circuit, then we may of course simply transfer the data back into the computational basis with a $H^\infty$ operation, instead of measuring it.

The aim in any rendition of Shor's Algorithm is not to consider such data in isolation but to chain together several of these $\Lambda_r(U_{\rm anc})$ gates, using a different register wire $r$ for each.  Note that these gates all commute, so there is no need to prespecify the order in which they are to be performed.  In this manner, many bits are provided for assisting in the estimation of $\theta$.

As it stands, this algorithm doesn't amount to much, because the classical bits issued by such a circuit are highly correlated and carry little information collectively.  Indeed, one would obtain a remarkably poor estimate of $\theta$ if this approach alone were relied upon. 
(It would require exponentially many samples of the $[\cos^2( \theta ) : \sin^2( \theta )]$ probability distribution to estimate $\omega$ to a sufficient accuracy.)
The real key to Kitaev's observation is to use other unitary maps that have the same eigenspaces.  These are the maps of the form $\Lambda_r(U^c_{\rm anc})$, and they all commute.  By chaining these maps into the computation, data can be collected not just on some specific $\omega$ eigenvalue, but on its powers, $\omega^c$, also.  This process can be extended to include as many commuting maps as we have patience and spare wires for, provided that it is no more difficult to find circuitry for implementing $\Lambda_r(U^c_{\rm anc})$.

\subsection{Ancilla in Order-Finding}
In practice, it will not be possible to load the ancilla register with a non-trivial eigenvector.  Instead, we load it with an arbitrary state.  Provided that this initial ancilla state does not overlap much with the eigenvalue-1 eigenspace, the effect of the Order-Finding part of the algorithm just described will be to generate within the input-register data that is in a superposition of states which (when understood in the Hadamard basis) estimate the different possible $\theta$ values.  To a great extent, it makes no difference whether the ancilla superposition is coherent or incoherent, so either an arbitrary pure state or a mixed state could be used.  Even a fully-mixed state can be used for the ancilla, since the eigenvalue-1 eigenspace doesn't dominate the `useful' spaces superpolynomially.

To make this more explicit, we show how to determine a stochastically chosen $\theta$ in the case where it is easy to find simple circuits for exponentially high powers of a certain $U$.
Starting with $a \cdot b$ input wires, each initialised as before to $\ket+$; for $j \in \set{ 1, 2,  \dots , a }$, for $c_j$ being an exponentially growing series of positive integers, we can reserve $b$ wires for each value of $j$, each for controlling one implementation of $U^{c_j}$.
We are free to choose any values we like for the $c_j$, but not all choices will be appropriate for completing the task of eigenvalue estimation.  More will be said about the $c_j$ values later.  
\beqa  \label{eqn:ShorOutput}
  \lefteqn{ \sum_{\omega} ~ \alpha_{\omega} \ket{\lambda_{\omega}}_{\rm anc} ~ \otimes ~ \ket+^{\otimes ab}_{\rm input} ~~~\mapsto } 
  \nonumber \\ 
  & & \sum_{\omega} ~ \alpha_{\omega} \ket{\lambda_{\omega}}_{\rm anc} ~ \otimes ~ \bigotimes_{j=1}^a
         \left( \frac{(1+\omega^{c_j})\ket+ + (1-\omega^{c_j})\ket-}{2} \right)^{\otimes b}.
\eeqa

The traditional rendering of Shor's algorithm proceeds by applying an appropriate Quantum Fourier Transform (QFT) to the input register at this stage \cite{NandC}, but since the QFT is not readily implementable using small numbers of the gates that we are wishing to consider, we shall instead consider this output state (\ref{eqn:ShorOutput}) to conclude the \emph{Order-Finding} part of our algorithm.

\subsection{Quantum Depth} \label{sect:Qdepth}
Sometimes it is convenient to think of the full algorithm as having quantum-depth of 1, allowing $\ket+$ states as input and where the two parts (order-finding and post-processing) are separated by a $H^\infty$ operator.  By performing a $H^\infty$ on the output state of (\ref{eqn:ShorOutput}) it is clear that the data in which we're interested are transformed from the $X$-basis to the $Z$-basis, where they can be processed with Toffoli gates.

Sometimes it is better to think of our algorithm as having quantum-depth 0, where the postprocessing part is performed not inside the circuit but on a classical computer \emph{after} the quantum algorithm has ended and the circuit output has been measured in the $X$-basis (assuming $X$-basis measurement is available.)  

Sometimes it makes more sense to think of our algorithm as having quantum-depth 2, for then we are free to use the canonical $\ket0$ input on each wire, and require no measurement in the $X$-basis at all.  

Therefore, rather than trying to answer definitively the question, ``What is the quantum-depth of our implementation of Shor's Algorithm?''  we shall instead proceed simply by describing how classical processing of samples from 
\beq  \label{eqn:probdist}
  \left\{~ x_j \sim \hbox{\emph{Binomial}}(~b, ~\sin^2(\theta \cdot c_j) ~) ~\right\}_{j=1}^a
\eeq
can be used to provide a good enough estimate of $\omega$, where $\omega$ is chosen at random with probability $|\alpha_{\omega}|^2$ (see expression (\ref{eqn:ShorOutput}),) and $\omega =: \exp( 2i\theta )$.  [To be clear, $x_j$ will be an integer between $0$ and $b$, most likely around about $b \cdot \sin^2(\theta \cdot c_j)$, \&c...]

\subsection{Parallelisation of Order Finding}
Before proceeding to explain the (classical) computational tasks involved in processing the output of \emph{Order Finding}, we make a few comments about the circuitry used so far.  Toffoli circuitry is used (as explained in subsection \ref{sect:OrderFinding}) to implement 
\beq
  \prod_j \Lambda_{R_j}( U_{\rm anc}^{c_j} ),
\eeq
where each $R_j$ is a set of $b$ wires used to control the application of the gate $U^{c_j}$, which is just a modular-multiplication mapping.  It is well known (\emph{e.g.} see \cite{Papadimitriou} for the case of integer multiplication) that circuitry for this kind of functionality can frequently be constructed to have overall \emph{polylog} depth in the input length, with a merely \emph{polynomial} overhead in the circuit width, paid for by addition of certain extra ancilla gates that are to be initialised to $\ket0$.  To provide such ancill\ae{} within our model would add a small constant factor to the overall quantum-depth of the algorithm, as well as increasing its width; but if we are prepared to pay such a cost, then the circuitry of \emph{Order Finding} can be said to be `exponentially parallelisable'.

\subsection{Eigenvalue Estimation}
Since having rejected the Quantum Fourier Transform as a way of recovering the desired (classical-description) approximation to $\omega$, we must instead address the \emph{purely classical} problem of figuring out how to choose the parameters $a$, $b$, and $c_j$, and how to use the samples $x_j$ in order to form the approximation to $\omega$ in an efficient manner.  
This is not an altogether straightforward task, because it depends on the problem being solved, \emph{e.g.} one chooses parameters slightly differently from the Integer Factorisation problem than for the Discrete Logarithm problem.

A little experimentation in the general case, implementing a sampling strategy for random $\theta$ in expression (\ref{eqn:probdist}) on a (classical!) computer, has led us to the conclusion that the optimal choice for $c_j$ is \emph{not} consistently $2^j$ as used with the QFT method, (see \cite{NandC}.)  This is because in the case where the binary expansion of $\theta$ has a long run of zeroes or ones after $j$ places, then there is remarkably little data in $x_j$.  

Experimentally, we found that if $\theta$ needs to be estimated to within one part in $K$, then it suffices (for a reasonable chance of estimating $\theta$ correctly) to take $b=3\log\log(K)$ and to let the $c_j$ take every value between $1$ and $K$ that either has just one bit set or else has two bits set which are not more than $b/2$ places apart.  Using those $c_j$ values with \emph{two} bits set enables one to write code which can successfully `bridge over' those region of the Real line (for $\theta$) that include numbers whose binary expansions have early long runs of zeroes or ones.  The precise details of this technique are omitted for brevity, but the problem is not technically difficult. 
Proving rigorous limits in this regime however remains an open problem.

In the case where $\theta$ is known {\it a priori} to be a multiple of $2\pi/\phi$ for some unknown integer $\phi \sim \sqrt{K}$, then this resolution suffices (with constant probability) to recover $\theta$ exactly, by the method of continued fractions.  There is extensive literature on applying Eigenvalue Estimation to specific problem-instances.

\subsection{Generalising ``Quantum-Depth = 2''}
We saw in the prequel that (a version of) Shor's Algorithm may be implemented in our computational model with overall quantum-depth of 2, starting from a trivial $Z$-basis state and ending with a $Z$-basis measurement.  
Consider now a `general' algorithm of quantum-depth 2, and the computational path it follows.
Starting from $Z$-basis state $\ket{a}$ (where $a \not= 0$) on $n$ wires; apply a $H^\infty$ map; then a $Z$-basis-permutation $T$; then another $H^\infty$; then measure the result $\ket{c}$ in the $Z$-basis.
This computation, and the resulting probability distribution, may be denoted by
\def\sgn#1#2{ (-1)^{ \left< #1,#2 \right> } }
\beqa
  \ket{a} &\mapsto& 2^{-n} \sum_{b,c} \sgn{a}{b} \sgn{T(b)}{c} \ket{c} \nonumber \\
          &\mapsto& \bigl\{~ \mbox{Prob}(c) ~=~ 2^{-n}\sum_b \sgn{a}{b} \sgn{T(b)}{c} ~\bigr\} \nonumber \\
          &=& \left\{~ \begin{array}{l}
                         \mbox{Prob}(0) ~=~ 0  \\
                         \mbox{Prob}(c) ~=~ 2^{1-n} \cdot \sum_{b:\left<a,b\right>=0} \sgn{T(b)}{c} 
                       \end{array}  ~\right\}.
\eeqa

It is somewhat surprising to reflect that this little formula contains virtually all the power needed to factorise large integers.
Note that as well as being implementable in the model this paper has described, algorithms with this low quantum-depth are also effectively implementable (modulo classical postprocessing) on a quantum computer that has access \emph{only} to gates of the form 
$~H^\infty \cdot \Lambda_C(X_T) \cdot H^\infty$, 
together with Z-basis inputs and outputs.  [Note that despite our notation, this gate acts trivially on all but a constant number of qubits, and so it is `small' in the usual sense.]  While such a machine could help us factorise large integers, it would be of little use for ordinary `classically easy' tasks.
The example of this ``conjugated Toffoli gate'' essentially resolves one of the open problems (problem 4) listed in \cite{Aaronson}~: The gate is neither classically simulable nor universal for quantum computation in any reasonable sense.  This assertion is justified by the observations~: 1) if it were classically simulable, then \emph{Order Finding} would be classically simulable too, and hence \emph{Factorisation} would be in {\bf BPP}; 2) if it were universal for {\bf BQP} then a polynomial usage of the conjugate-Toffoli could approximate one usage of an ordinary Toffoli gate, so by symmetry a polynomial usage of the Toffoli gate could approximate one conjugate-Toffoli gate, and hence the latter would be classically simulable. 
To find uses for such a ``conjugate-Toffoli machine'', beyond applications of Shor's Algorithm, we leave as an open problem.

%% SECTION FIVE :  Grover
\section{Grover's Algorithm}  \label{sect:Grover}
\subsection{Simple Oracle for the Unstructured Search Problem}
Define a \emph{simple Grover Oracle} $G_x$, with \emph{hidden data} $x$, to be an application of the following one-register phase-flip map.  (Note that the Grover Oracle does not return a description of a circuit for implementing that mapping, rather it simply returns an application of that mapping, as per the `black-box' model.)
\beq
  G_x \ket{z} ~:=~ (-1)^{ \indicator{x=f(z)} }\ket{z}.
\eeq
Here $f$ is taken to be a fixed function acting on the labels of the computational basis of the register, and $\indicator{x=f(z)}$ is $1$ if $x=f(z)$ and zero otherwise.  We let $N$ denote the cardinality of the range of $f$.  It is intended that $f$ have very little complexity~: typically it will just check the first $\log(N)$ qubits of the register.

A simple Grover Oracle for an unstructured \emph{search problem} may be simulated in the model pertinent to our present study, provided that there is an efficient way of determining classically whether $x=f(z)$ for any given $z$.  
For example, this can be managed with a two-qubit ancilla in the state $\ket{1-}$, so that no matter how many times $H^\infty$ has been applied, there will still be a qubit in the state $\ket-$ on which to target a generalised Toffoli gate, so as to implement the necessary sign-flip that the oracle (simulation) requires.

\subsection{Search Algorithm Notation}  \label{sect:SearchAlgNotate}
A \emph{Grover Search Algorithm} expressed in our model comprises a circuit of $\Lambda_C(X_T)$ gates and $H^\infty$ gates and an input, and a computational basis measurement at the output which, with high probability, will result in a $\ket{z}$ that indicates the hidden data $x=f(z)$.  We are interested in lower-bounding not the depth of the circuit but rather its quantum-depth and the number of oracle calls used.  The notation for the algorithm is 
\beq
  G_x^{k_T} \cdot H^\infty \dots H^\infty \cdot G_x^{k_2} \cdot H^\infty \cdot G_x^{k_1} \cdot \ket{\psi},
  \nonumber 
\eeq
where
\beq  \label{eqn:GroverAlg}
  G_x^{k_t} ~:=~ T_{t,k_t} \cdot G_x \dots T_{t,2} \cdot G_x \cdot T_{t,1} \cdot G_x \cdot T_{t,0},
\eeq
where $T_{t,j}$ are predetermined permutations of the $Z$ basis.
The quantum-depth of this circuit is $T-1$ and the number of oracle calls is $\sum_{t=1}^T k_t$.

It would be trivial to specify how to implement Grover's Algorithm if we didn't care about quantum-depth; simply interleave calls to the oracle with maps of the form $H^\infty \cdot T_0 \cdot H^\infty$, for some appropriate $T_0$.  But the questions we should like to ask concern lower-bounds and trade-offs between the quantum-depth and the number of oracle calls required.

\subsection{Categories of Success}
To be precise, we must state for which kinds of algorithm are we asking for bounds.  Consider deterministic algorithms, algorithms that have worst-case probability above some bound, and algorithms that have average-case probability above some bound.

Deterministic algorithms are not the focus of this study.  While of theoretic interest, such algorithms are less amenable to the kinds of analysis that we wish to deploy, and the model described in this article probably has little to contribute to their study.  

Algorithms with \emph{bounded worst-case probability} are those for which, no matter what the hidden data turns out to be, the probability of the algorithm succeeding exceeds $1-\epsilon$ for some constant $\epsilon$.  Since the success probability of an algorithm can always be `amplified' by running it several times (either in series or in parallel), without loss of generality we always take $\epsilon$ to be less than $1/2$.  This kind of algorithm is probably the easiest to bound, (see \cite{0002066} for a good general technique for non-geometric proofs in this area,) but they shall not be of primary concern to us here.

Algorithms with \emph{bounded average-case probability} are the most relevant for an algorithm designer, since they allow for the possibility of breaking some symmetry in the problem in such a way that certain answers will be substantially easier to recover.  They are algorithms for which \emph{a priori} the success probability exceeds $1-\epsilon$.  

For highly symmetric problems (such as \emph{Unstructured Search}) one generally expects such algorithms to be the same as those with bounded worst-case probability, but the techniques for proving average-case bounds can sometimes be different.  Given many instances of a certain kind of `naturally arising' problem, an algorithm with bounded average-case probability will be most welcome, since it will almost certainly serve to address many of those instances.  It is less use if the instances in hand are actually derived from a different kind of problem that just happens to lie `nearby' in terms of complexity, since there might be no reason to assume that what is `average' for one class of problem will translate into average instances as derived from that class.

Henceforth we will consider only lower bounds for the resource requirements of algorithms with bounded average-case probability, since a lower bound for resource requirements in the average case implies a lower bound for resource requirements in the worst case, but the converse does not follow.

\subsection{Lower bounds}
In this section we give some preliminary lemmas leading to a proof of the following~:

\begin{theorem}  \label{thm:Grover}
  Any Grover Search Algorithm, as defined in section \ref{sect:SearchAlgNotate} with the notation of equation (\ref{eqn:GroverAlg}), with some constant lower bound for its average-case expected probability of success, will require sufficient oracle calls between $H^\infty$ stages so that 
  \beq
    \sum_{t=1}^T \sqrt{k_t} ~=~ \Omega( \sqrt{N} )
  \eeq
  holds true asymptotically.
\end{theorem}
Note that this means an algorithm might make one oracle call in each of $\sim \sqrt{N}$ quantum-depth-phases, or it might make $\sim N$ calls without any quantum-depth, or it might interpolate these extremes.

%% LEMMA 1
\begin{lemma}  \label{lemma:magic}
  Let $w_{x,t}$ be entries of a 2-dimensional array of non-negative real numbers.
  Let $R_t = \sqrt{\sum_x w_{x,t}^2}$ be the 2-norm of the rows, and
  let $C_x = \sum_t w_{x,t}$ be the 1-norm of the columns.
  Then let $R = \sum_t R_t$ be the 1-norm of the $R$s,
  and  let $C = \sqrt{\sum_x C_x^2}$ be the 2-norm of the $C$s.
  Then $R \ge C$.
\end{lemma}
To prove this, let $t$ be the vector whose $x$th entry is $w_{x,t}$.  Then using the Euclidean inner product, we can rewrite $R_t = \sqrt{\left<t,t\right>}$.  By expanding the expression for $C^2$ we get $C^2 = \sum_{s,t} \left<s,t\right>$.  Likewise, $R^2 = \left( \sum_t \sqrt{\left<t,t\right>} \right)^2 = \sum_{s,t}\sqrt{\left<s,s\right> \cdot \left<t,t\right>}$.
So it suffices to show that always
\beq
  \sqrt{\left<s,s\right> \cdot \left<t,t\right>} ~\ge~ \left<s,t\right>,
\eeq
but this is a basic (Euclidean) trigonometric inequality.
\QED

%% LEMMA 2
\begin{lemma}  \label{lemma:sphericaltriangle}
  Given three unit vectors in a Euclidean Geometry, the sum of the absolute values of the sines of any two inter-angles is no less than the absolute value of the sine of the third inter-angle.
\end{lemma} 
This is trivial in two dimensions.  The general problem is three-dimensional.
Consider three unit vectors whose components are $(1,0,0)$, $(a,b,0)$, and $(c,d,e)$, with $|b|\ge0$.  (No generality is lost with this choice.)  Then it only remains to show that 
\beq
  b + \sqrt{1-c^2} ~\ge~ \sqrt{ 1 - (ac+bd)^2 }.
\eeq
We begin by squaring both sides, so that we are required to show that
\beq
  b^2 + 1 - c^2 + 2b\sqrt{1-c^2} ~\ge~ 1 - (ac+bd)^2.
  \label{eqn:strange}
\eeq
Next, we try to balance between $d$ and $e$.  The case $d=0$ immediately satisfies (\ref{eqn:strange}).  The case $e=0$ reduces (\ref{eqn:strange}) to the requirement $|ac+bd| \le 1$, which is a basic trigonometric result, as the reader can easily check.  The only remaining case occurs when the partial derivative of (\ref{eqn:strange}) w.r.t. $d$ vanishes, \emph{i.e.} when $(ac+bd)b = 0$.  The $b=0$ sub-case is easily dismissed, and the other sub-case is defined by $ac=-bd$.  We can use this to reduce (\ref{eqn:strange}) to $b^2-c^2+2b\sqrt{1-c^2}\ge0$, which by solving for $b$ is identical with the condition $b\ge1-\sqrt{1-c^2}$ on the range of validity.  We can also use it to obtain $(1-b^2)c^2 = b^2d^2$, and hence $|c|\le b$.  And this latter inequality guarantees the condition sought~: $\sqrt{1-c^2} \ge \sqrt{1-b^2} \ge 1-b$.  \QED

%% LEMMA 3
\begin{lemma}  \label{lemma:sumofReals}
  For $(t_x)$ a list of real numbers, 
\beq
  \sum_{x=1}^N t_x^2 ~\le~ Np \quad\Rightarrow\quad \sum_{x=1}^N t_x ~\le~ N\sqrt{p}.
\eeq
\end{lemma}
The variance of $t_x$ is non-negative, that is
\beq
  \frac1N \sum_x t_x^2 ~-~ \left( \frac1N \sum_x t_x \right)^2 ~\ge~ 0,
\eeq
and the lemma follows directly from this observation.   \QED
\bigskip

%% PROOF OF THEOREM
To prove theorem \ref{thm:Grover}, we proceed as follows.
Referring back to the notation of section \ref{sect:SearchAlgNotate}, let
\beqa 
  U_t &:=& H^\infty \cdot T_{t,k_t} \dots T_{t,0};  \nonumber \\
  \ket{\psi_t} &:=& U_t \cdot U_{t-1} \dots U_1 \cdot \ket{\psi};  \nonumber \\
  P_{x,t} &:=& U_t^\dag \cdot H^\infty \cdot G_x^{k_t}~; \nonumber \\
  \ket{\psi_{x,v,t}} &:=& U_v \cdot P_{x,v} \cdot U_{v-1} \cdot P_{x,v-1} \dots U_{t+1} \cdot P_{x,t+1} \cdot \ket{\psi_t}; 
  \label{eqn:psiXVT} 
\eeqa
and write
\beqa
  {\rm path}_{x,t}(z) &:=& \indicator{f \cdot T_{t,0}(z)=x} \oplus \dots \oplus \indicator{f \cdot T_{t,k_t-1} \dots T_{t,0}(z)=x}; 
    \nonumber \\
  W_{x,t}(\ket{\phi}) &:=& \sum_z \bigl| \braket{\phi}{z} \bigr|^2 \indicator{ {\rm path}_{x,t}(z) \equiv 1 } ~~(\le1);
    \label{eqn:W} 
\eeqa
Then it follows that
\beq  \label{def:P} 
  P_{x,t}\ket{z} ~=~ (-1)^{{\rm path}_{x,t}(z)}\ket{z};
\eeq
and states defined in (\ref{eqn:psiXVT}) may be compared with each other according to
\beqa
  \braket{\psi_{x,T,t}}{\psi_{x,T,t-1}} &=& \bra{\psi_{t-1}} \cdot P_{x,t} \cdot \ket{\psi_{t-1}} 
    \nonumber \\
    &=& 1 - 2W_{x,t}(\ket{\psi_{t-1}}), \nonumber \\
  1 - \bigl| \braket{\psi_{x,T,t}}{\psi_{x,T,t-1}} \bigr|^2 &\le& 4W_{x,t}(\ket{\psi_{t-1}}).
    \label{util:W} 
\eeqa
Square-rooting expression (\ref{util:W}), summing that over the $t$ values, and further approximating by many applications of lemma \ref{lemma:sphericaltriangle}, then squaring, we obtain 
\beq  \label{eqn:approx1}
  1 - \bigl| \braket{\psi_T}{\psi_{x,T,0}} \bigr|^2 
    ~\le~ 4\left( \sum_{t=1}^T \sqrt{W_{x,t}(\ket{\psi_{t-1}})} \right)^2.
\eeq 
Using the trigonometric notation
\beq  \label{indepx}
  \cos( \ket{\psi_T}, \ket{\psi_{x,T,0}} ) ~=~ \bigl| \braket{\psi_T}{\psi_{x,T,0}} \bigr| 
\eeq
and summing expression (\ref{eqn:approx1}) over the $x$ values, we obtain,
\beq  \label{eqn:approx2}
  \sum_x \sin^2(\ket{\psi_T}, \ket{\psi_{x,T,0}})
    ~\le~ 4\sum_x\left( \sum_{t=1}^T \sqrt{W_{x,t}(\ket{\psi_{t-1}})} \right)^2.
\eeq 
Denote by $F$ the left side of this inequality.  Apply lemma \ref{lemma:magic} to the right side of this inequality after square-rooting, and it yields
\beq  \label{eqn:approx3}
  \sqrt{ F }
    ~\le~ 2\sum_{t=1}^T \sqrt{\sum_x W_{x,t}(\ket{\psi_{t-1}})}.
\eeq 
Now, by equation (\ref{eqn:W}), we know that
\beqa 
  \sum_x W_{x,t}(\ket{\psi_{t-1}}) 
  &=& \sum_z \bigl| \braket{\psi_{t-1}}{z} \bigr|^2 \sum_x \indicator{ {\rm path}_{x,t}(z) \equiv 1 }
    \nonumber \\
  &\le& \sum_z \bigl| \braket{\psi_{t-1}}{z} \bigr|^2  \cdot k_t \nonumber \\
  &=& k_t.
    \label{eqn:approx4}
\eeqa 
So if we put these observations together (substitute (\ref{eqn:approx4}) into (\ref{eqn:approx3})), we see
\beq 
  F ~\le~ 4\left( \sum_{t=1}^T \sqrt{k_t} \right)^2.
\eeq 
It suffices to show that this $F$ is lower-bounded by $\Omega(N)$ whenever the algorithm has expected success-probability $\Theta(1)$.
Let $\ket{C_x}$ denote the closest unit vector to $\ket{\psi_{x,T,0}}$ that is in the span of $\{ \ket{z} : f(z) = x \}$.
Since the output of the algorithm is $H^\infty \cdot \ket{\psi_{x,T,0}}$, we can use a measurement in the basis given by $H^\infty \ket{C_x}$, and by the Born rule establish that the expected probability of success is  
\beq
  p ~=~ N^{-1}\sum_x \cos^2( \ket{\psi_{x,T,0}}, \ket{C_x} ),
\eeq
which we rewrite as 
\beq
  \sum_x \sin^2( \ket{\psi_{x,T,0}}, \ket{C_x} ) ~=~ N(1-p).
\eeq
It follows from lemma \ref{lemma:sumofReals} that 
\beq  \label{eqn:approx5}
  \sum_x |\sin( \ket{\psi_{x,T,0}}, \ket{C_x} )| ~\le~ N\sqrt{1-p}.
\eeq
And since the $\ket{C_x}$ are all mutually orthogonal, $\sum_x \cos^2( \ket{\psi_T}, \ket{C_x} ) \le 1$, 
which we write as
\beq  \label{eqn:approx6}
  \sum_x \sin^2( \ket{\psi_T},\ket{C_x} ) ~\ge~ N-1.
\eeq
Now apply lemma \ref{lemma:sumofReals} directly to the definition of $F$ (LHS of inequality (\ref{eqn:approx2})) to obtain
\beq  \label{eqn:approx7}
  \sum_x |\sin( \ket{\psi_T}, \ket{\psi_{x,T,0}} )| ~\le~ \sqrt{N \cdot F}.
\eeq
Putting together lines (\ref{eqn:approx5}), (\ref{eqn:approx6}), (\ref{eqn:approx7}), and applying lemma \ref{lemma:sphericaltriangle}, we obtain 
\beq
  \sqrt{N \cdot F} + N\sqrt{1-p} ~\ge~ N-1,
\eeq
whence $p = \Theta(1) ~\Rightarrow~ F = \Omega(N)$, as required.  \QED
\medskip

%% CONCLUSIONS
\section{Conclusions}
We have introduced a new theoretical model for quantum circuits, designed to highlight one aspect of the way in which quantum computation differs from classical computation.  We have thence illustrated a little of what can be achieved within limited quantum-depth, by analysis of the two main (or well-known) algorithms of Quantum Information Processing; showing that `hard' classical problems can sometimes be `solved quantumly' using only Toffoli gates, and using the model to add to the growing literature on ``algorithmic trade-offs''.  We have developed a few tools for facilitating these analyses, and exemplified quantum a gate that is probably neither \textbf{BQP} universal nor classically simulable.  Further exploration of the power of computing within very small (\emph{e.g.} constant) quantum depth remains an interesting issue for future research.

\section*{Acknowledgements}
Thanks are due to Richard Jozsa for much proof-reading and helpful discussions.  This work has been sponsored by CESG.

\end{document}